# Insights from the Wikipedia Contest

*Kalpit V Desai, Roopesh Ranjan*

## Abstract

The Wikimedia Foundation has recently observed that newly joining editors on Wikipedia are increasingly failing to integrate into the Wikipedia editors' community, i.e. the community is becoming increasingly harder to penetrate [1]. To sustain healthy growth of the community, the Wikimedia Foundation aims to quantitatively understand the factors that determine the editing behavior, and explain why most new editors become inactive soon after joining. As a step towards this broader goal, the Wikimedia foundation sponsored the ICDM (IEEE International Conference for Data Mining) contest [2] for the year 2011.

The objective for the participants was to develop models to predict the number of edits that an editor will make in future five months based on the editing history of the editor. Here we describe the approach we followed for developing predictive models towards this goal, the results that we obtained and the modeling insights that we gained from this exercise. In addition, towards the broader goal of Wikimedia Foundation, we also summarize the factors that emerged during our model building exercise as powerful predictors of future editing activity.

## Data

The training dataset contained edit history on the English Wikipedia from the period January $1^{st}$, 2001 - August $31^{st}$ 2010. The edit history was provided for a random sample of 44514 editors out of all those editors who had made at least one edit in the year ending in August $31^{st}$ 2010. The edit history, disclosed as a part of the contest dataset, contained following information for each edit that was made:

- Masked ID of the editor who made the edit, which can be mapped to the date the editor registered into Wikipedia community
- The type (namespace) of the edit e.g. main article, discussion, user, etc.
- ID of the article being edited, which can be mapped to article title, article category (e.g. list, featured article, feature picture etc.) and article creation date
- Any text comments that the editor may choose to put describing the edit
- time stamp and revision number of the Wikipedia upon edit
- if the edit got reverted, then ID of the user who reverted it and the reverted-to revision#
- Net edit size in characters, article size in characters after the edit, and MD5 hash of the article after the edit.

The dataset didn't contain edit history of anonymous editors, or on deleted articles.

## Evaluation Metric

The Root Mean Squared Logarithmic Error ("RMSLE") was used to measure the accuracy of an algorithm. The RMSLE is calculated as

$$\epsilon = \sqrt{\frac{1}{n}\sum_{i=1}^{n}\left(\ln\left(\frac{p_i + 1}{a_i + 1}\right)\right)^2}$$

$\epsilon$ is the RMSLE value (score)
$n$ is the total number of editors in the data set
$p_i$ is the predicted number of edits made by editor $i$ during Sep 01, 2010 to Jan 31, 2011
$a_i$ is the actual number of edits made by editor $i$ in the 5 month period

## Preliminary Observations

A. **Number of edits made by an editor in a given period follows Pareto distribution.**

The scaling or Pareto Distribution [3] is characterized by the equation $\Pr(X > x) = x^\lambda$. For $\lambda$ = 1.161, the Pareto distribution becomes the classic 80-20 law i.e. 80% of all income is received by 20% of all people, and 20% of the most affluent 20% receive 80% of that 80% income and so on. As depicted in figure 1, when fitted against the distribution of number of edits by editors, we found that $\lambda$ = 2.51 describes this scaling distribution reasonably well. Logarithm of a standard scaling distribution follows an exponential distribution.

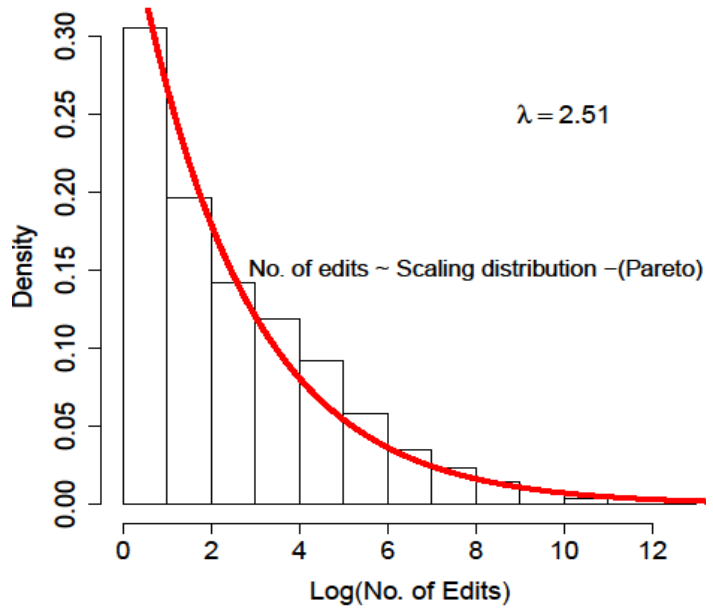

Figure 1: Number of edits as Pareto distribution

B. **Survivorship Bias:** The organizers of the contest deliberately excluded from the contest dataset those editors who made no edits during the year ending in Sep 1$^{st}$ 2010. Their rationale [4] behind the chosen over-representation of active editors is that if chosen the alternative, i.e. a random sample of the editors, more than 80% would be inactive. Building models that achieve high accuracy on this sample of mostly idle editors will not help the Wikimedia Foundation understand the editing behavior.

An Interesting side-effect of this sampling is a survivorship bias that affects only those editors who joined before September 2009. As we show below, more than half of the editors in the dataset have joined after that date; thus aren't affected by the survivorship bias. This disparate sampling effect motivated us to segment the editors based on their joining date, and for each type of models, fit separately for each segment.

C. **More than half of the active editors are less than 1 year old in the system**

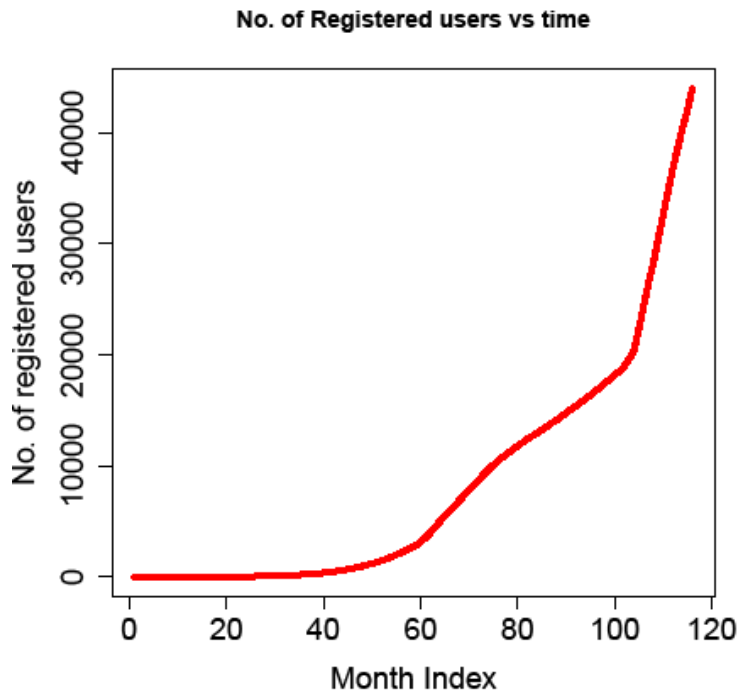

Figure 2: Cumulative count of registered editors

Figure 2 shows the number of registered editors versus number of months elapsed since Jan 2001. As seen, approximately 24000 editors, or 55% of all editors in the contest database, have joined in the last year (month index 105 to 116). This overwhelming proportion of relatively new editors for whom we only have less than a year of editing history makes the prediction task even more challenging.

D. **Over-all edit rate is decaying with time, and individual edit Rate is a strong function of the editor's age in the system.**

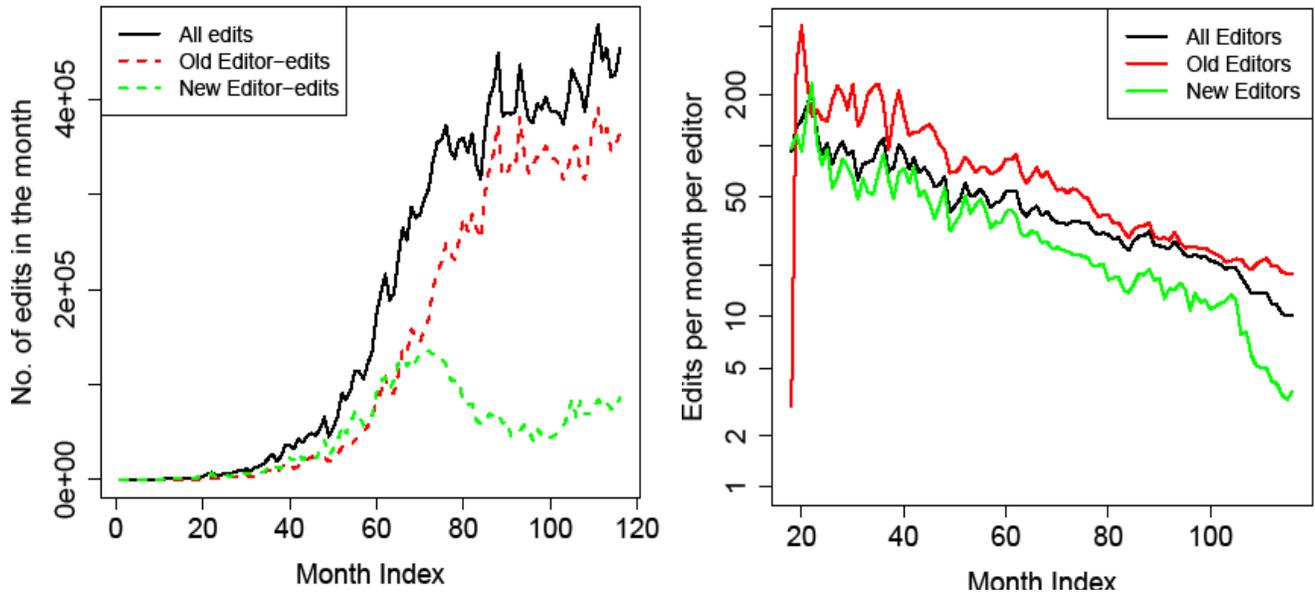

Figure 3a (left): Monthly total edits for each group of editors. 3b (right) Monthly averaged (per-editor) edits for each groups of editors.

As illustrated by Figures 3a, majority of the edits are contributed by the old editors (more than 1 year in the system). Also as shown in Figure 3b, per-editor monthly edit rate is significantly higher for old editors as compared to new editors. Also for both groups of editors, the edit rate is decreasing with time.

Additionally, several insightful trends have been reported in this document from Wikimedia Foundation [1] and this paper [5].

## Our Approach

The table 1 summarizes the significant milestones along our journey. We approached the problem by starting with the simplest model, i.e. the persistence model where for each editor the prediction for the future five months is simply the number of edits that the editor made in the last five months (i.e. the *persistence period*). We further enhanced this model by leveraging the insight that the overall edit rates are declining; we optimized a persistence downscaling factor α against RMSLE. We also explored linear models by adding a few more features to the downscaled persistence model, e.g. number of edits made in last month, number of edits made in the five months period immediately before persistence period, no of days in last five month on which edit was made etc. The last section of the article describes the features we found most useful for predicting future edits. We obtained the values of the coefficients for these linear models by using the non-linear optimization routine *nlm* available in R [9] to minimize RMSLE. This choice of using non-linear optimizer (instead of doing linear least squares etc) was motivated by the nonlinear nature of the RMSLE loss function.

Based on the previously gained insights about the data, we also explored two ways of segmenting editors: by their age into the system, and by their level of activity in the persistence period. Both of these segmentation approaches produced significant improvements. However, any further segmentation of editors didn't improve the accuracy. At the extreme, fitting a separate model for each editor, all types of models we tried underperformed the simple downscaled persistence model.

We also found, to our surprise initially, that linear ensembles of models produced prediction accuracy that was often worse than all of the constituent models of the ensemble. Similarly Bootstrap aggregated models [6] also underperformed the models created with simple 80/20 splits. The reason, as we realized later, was the default method of aggregation i.e., arithmetic aggregation. The arithmetic aggregation is not appropriate for a random variable that follows a scaling distribution. Once we leveraged this insight and created ensembles of models using geometric aggregation, the prediction accuracy of the ensemble significantly improved.

Other sophisticated machine-learning techniques such as Artificial Neural Networks and Support Vector Machines turned out to underperform. We suspect this was due to the difficulty in integrating and leveraging the available insights about the data into those sophisticated models. However, using Random Forest models, after a careful feature selection and appropriate segmentation of editors, we were able to obtain accuracy comparable to our best models.

|         | Model Description | Test RMSLE |
|---------|-------------------|------------|
|         | Persistence Model | 1.129406 |
|         | Persistence with optimized downscaling<br>No segments (1 parameter) | 1.009079 |
| Model P | Persistence with optimized downscaling<br>3 segments (3 parameters) | 0.957141 |
| Model Q | Model P with optimized Intercept<br>3 segments (6 parameters) | 0.922552 |
|         | Random Forest models<br>2 segments<br>(14 / 19 features, 150 / 200 trees) | 0.910002 |
|         | Linear model with nested segments<br>7 segments (43 parameters) | 0.909674 |
| Model R | Model Q with age in system, and edit days in persistence period<br>3 segments (12 parameters) | 0.905326 |
|         | A linear model for the residuals of Model R<br>3 segments (12 + 6 parameters) | 0.895141 |
|         | Model R with interaction terms<br>3 segments (24 parameters) | 0.884733 |
|         | Ensemble of 8 models using geometric aggregation (winner of Honorable Mention). | **0.869071** |

Table 1: Selected milestones along the model-building journey

**Final Algorithm**

*Following models went into the ensemble that produced our best performing entry with Root Mean Squared Log Error (RMSLE) of 0.869071 on the whole test set. This entry won the Honorable Mention Prize in the contest.*

A. **Log-Log Model (Model-1):** This model used more than 80 features, most of which were a linear function of no. of edits in various time spans. No. of edits being a scaling distribution we decided to use the log-transform to make it stable. Here we fitted a linear model for log of no. of edits) on the log of features. Below we give the form of the model where, $x_1, x_2, x_3, \ldots$ are features used in prediction and $\beta_1, \beta_2, \beta_3, \ldots$ are the coefficients to be optimized. The coefficients were obtained by fitting last 5 month's edits as a function of the features by optimizing root mean square error (RMSE). As the model is fitted on the logarithmic scale this model indirectly also minimizes the RMSLE loss function. This model gave a RMSLE of 0.911 on the full test data.
Please note that $y$ = number of edits by the editor in the 5 months (training period).

$$log(y + 1) = \beta_0 + \beta_1 log(x_1) + \beta_2 log(x_2) + \cdots \qquad (1)$$

B. **Linear Model with Segments by Joining Date (Model-2):** As seen above editors who join late have lower edit rate as compared to the older editors. To take this into account we segment the editors into three groups namely (a) Editors who joined in the last 5 months (b) Editors who joined between 5-12 months and (c) Editors who joined more than 12 months ago. The model coefficients were obtained by optimizing RMSLE. This model scored 0.905 on the full test-set. The form of the model fitted separately for each segment is given below.

$$y = \beta_0 + \beta_1 x_1 + \beta_2 x_2 + \beta_3 x_3 + \cdots \qquad (2)$$

C. **Linear Models with Interactions (Models 3-5):** These models were built by adding interaction terms in the Model-2. Two interaction features were added to generate 3 different interaction models for each segment. The score for these models is in the range of 0.870 to 0.877. These were our best individual models. The form of these models is given below.

$$y = \beta_0 + \beta_1 x_1 + \beta_2 x_2 + \beta_3 x_3 + \gamma_1 x_1 x_2 + \gamma_2 x_2 x_3 + \cdots \qquad (3)$$

D. **Linear models with Nested Segmentation (Models 6 -7):** This model applied two levels of segmentations on users. First level had two segments based on the registration date - those who have been in the system for at least 1 year, versus those who are newer. These two segments were further divided by the second level segmentation based on the number of unique edit days that the user was active in the persistence period. For each of these segments a separate linear model was fitted on chosen features (different for each segment), by running nonlinear optimizer to minimize RMSLE. The forms of these models were same as that of model B (2). The features were selected by manual forward selection method. Also, for each segment, we obtained 25 models by drawing 25 bootstrapped samples from the training set, and then aggregated the predictions by taking median. Another variant used a different set of features and aggregated using geometric mean. These models gave test scores of 0.909 and 0.911 respectively.

E.  **Random Forest Model (Model-8):** This model trained a separate Random Forest [7] Model for people who have joined before 2009-09-01 (i.e. old editors), versus those who have joined after that time (new editors). This partition was motivated by the survivorship bias that exists as we mentioned above. The RF model for new editors used 14 variables, whereas the RF model used for old editors used 19 variables, each determined by manual forward selection process. The model gave a test score of 0.91.

# Insights Gained

## Begin Simple and "Listen" to the Data

From the data-mining point of view, here is one of the most important insights we gained from this experience. Instead of approaching a data-mining problem with a set of standard automatized tools, interactive explorations of the data ("asking" and "listening") reveal very powerful insights upfront. A simple yet insight-driven model is usually far more effective than a very sophisticated and powerful model that doesn't adequately leverage the insights from the data. This observation isn't new for the data mining community; in fact it is perhaps the most popular rule of thumb in the field. Yet it is also routinely ignored as suggested by the fact that early half of the entries on the contest's leader-board [8] underperform the simplest model, i.e. the persistence benchmark. Also, compared to sophisticated models, simple models are far more amenable to tracing back the source of unexpected or surprising outputs of the model. This diagnostic exercise often turns out to be precisely the path that leads to a hidden insight about the data.

## Understanding the Editing Behavior on Wikipedia

Also, following insights emerged during our model building exercise that may be potentially useful to understand the future editing behavior on Wikipedia. We observed that the overall edit rate averaged across all editors is decreasing. Consequently, a <u>downscaled persistence model</u> which multiplies past edits by a scaling fraction went a long way in predicting future edits. In the models that we fitted, the optimal downscaling factor for the new editors came out to be much smaller than that for the old editors – thus supporting the Wikimedia's observation that new editors' editing rate is decaying more quickly compared to that of old editors.

We also observed that if the editor has often edited in the recent past, then he is likely to be more active in the immediate future. How often (days/sessions) the editor has edited in recent past is a better predictor than how many edits that he has made in the recent past.

The <u>age of an editor in Wikipedia</u> emerged as a significant predictor of future edits, and an effective way to group the editors. <u>How often an editor was active in recent past</u> (e.g. five months) also turned out to be a powerful way for grouping. For a subset of editors that have been inactive (zero edits) in last five months, <u>how often they edited in last six-to-ten months</u> came out to be the most powerful predictor by far. For new editors the <u>time of last edit</u> was also an important predictor. <u>Number of reverts gotten</u>, which may capture the resistance experienced by the editor, was an important predictor for old editors who edited on more than two days in last five months; whereas the <u>number of reverts made</u> was an important predictor for new editors who edited on more than two days in last five months.

The interaction between number of edits in the persistence period and number of edits in the five months immediately before the persistence period was also a key predictor for future edits. Another important interaction was between <u>number of edits by the editor in last month</u>, and number of unique days of edits by the editor in the persistence period. These interaction parameters may capture the <u>consistency of an editor's editing activity</u> over time.

# Conclusion

Our experience of participating in this contest was a very educational and rewarding journey. We gained several insights from modeling philosophy perspective as well as from the perspective of understanding the important predictors for future editing behavior of Wikipedia editors. We hope that the insights about how the editors can be grouped and what factors determine the future editing activity of different groups will help the Wikimedia Foundation towards their broader objective of sustaining healthy growth of the Wikimedia community.

# Acknowledgments

We are sincerely thankful to our colleagues for their enthusiasm towards the contest and insightful discussions along the way. Specifically, we would like note that Ramasubramanian Sundararajan, Vignesh T.S., Manisha Srivastava, Angshuman Saha, Jayanth Marasanapalle and Debasis Bal have made significant contributions and provided support throughout our journey.